\documentclass[10pt,
showpacs,
preprintnumbers,amsmath,amssymb,
aps,prd,nofootinbib,eqsecnum,a4paper]{revtex4}
\usepackage{epsfig}
\usepackage{graphicx,epsf}
\usepackage{color}
\usepackage{bm}
\usepackage{psfrag}
\usepackage[symbol]{footmisc}
\usepackage{tensor}
\def\be{\begin{equation}}
\def\ee{\end{equation}}
\def\bea{\begin{eqnarray}}
\def\eea{\end{eqnarray}}
\def\bi{\begin{itemize}}
\def\ei{\end{itemize}}
\def\p{\partial}

\def\H{{\cal H}}

\def\cs2{c_{\rm{s}}^2}
\def \beg {\begin{enumerate}}
\def \en {\end{enumerate}}

\def\M0{{\cal M}_0}

\def\G{\mathcal{G}}
\def\H{\mathcal{H}}

\begin{document}

\title{Linear cosmological perturbations in almost scale-invariant
  fourth-order gravity}

\author{Jorge L.~Fuentes$^{1}$ \footnote[3]{\href{mailto:j.fuentesvenegas@qmul.ac.uk}{j.fuentesvenegas@qmul.ac.uk}} }
\author{Usman A. Gilani$^{1,2}$ \footnote[2]{\href{mailto:ugilani@student.qau.edu.pk}{ugilani@student.qau.edu.pk}}}
\author{Karim A.~Malik$^{1}$ \footnote[1]{\href{mailto:k.malik@qmul.ac.uk}{k.malik@qmul.ac.uk}}}

%
\affiliation{
$^1$Astronomy Unit, School of Physics and Astronomy, Queen Mary University of London, Mile End Road, London, E1 4NS, United Kingdom\\
$^2$Department of Mathematics, Quaid-i-Azam University, Islamabad, Pakistan
}
\date{\today}

\begin{abstract}
We study a class of almost scale-invariant modified gravity theory,
using a particular form of $f(R,G)= \alpha R^{2}+\beta G \log G$ where
$R$ and $G$ are the Ricci and Gauss-Bonnet scalars, respectively and
$\alpha,\beta$ are arbitrary constants. We derive the Einstein-like
field equations to first order in cosmological perturbation theory in
longitudinal gauge.
\end{abstract}

\pacs{98.80.Cq \hfill  arXiv:1812.XXXX}

\maketitle


\section{Introduction}
\label{intro}

Recent observations show that the universe is expanding at an
accelerated rate, at the moment the cause of this is still unknown. In
the cosmological standard model, one can assume different mechanisms
to describe dark energy (DE) such as the cosmological constant
$\Lambda$, or scalar fields: quintessence, $k$-essence, and many other
alternatives \cite{b,c,LDS2007}. Current surveys are being planned
like DES \cite{DES}, DESI \cite{Guy:2016zel}, Euclid
\cite{Tereno:2015hja}, and LSST \cite{LSST} to probe large scales in
order to find an answer to this problem.

Dark energy in general relativity (GR) is usually considered as a
change in the energy momentum tensor, $T_{\mu \nu}$, however one can
change the left hand side of the Einstein field equations and take the
accelerated expansion as an effect coming from the geometry of
spacetime, this is usually called modified gravity. Research has
focused until recently in models with $f(R)$, a function of the Ricci
scalar \cite{f,8f,PJ2007,CLF2010,SS2003}, but one can also focus on
more complex models like $f(R,T)$, where $T$ is the trace of the
energy momentum tensor \cite{TFS2011}. Among this modified theories of
gravity, Gauss-Bonnet (GB) gravity has been widely studied in its
$f(G)$ approach \cite{SSD2009,ERV2010}, where gravity is required to
couple with some scalar field $G$ \cite{SS2011,BI2007,SSS2002}. Recent
work \cite{HD2013} has focused on the study of ``almost
scale-invariant theories'' where $f(R,G)$ is a function of the Ricci
scalar and the GB term
\begin{equation}
\label{eq:gb}
G=R^2-4R_{\mu\nu}R^{\mu\nu}+R_{\mu\nu\rho\sigma}R^{\mu\nu\rho\sigma}\,,
\end{equation}
where $R_{\mu \nu}$ is the Ricci tensor and $R_{\mu \nu \rho \sigma}$
is the Riemann tensor. This model was initially proposed as a
gravitational alternative for DE and inflation in Ref.~\cite{g} and
its applications to late-time cosmology have been studied in
\cite{g,g1,g2}.

Modified theories of gravity do have their problems, such as
Ostrogradsky instabilities \cite{Becker:2017tcx,Motohashi:2014opa} and
ghosts \cite{Himmetoglu:2009qi}, but with the particular choice of the
a Lagrangian one can avoid this typical problems and find cosmological
solutions as power-law inflation and local attractors
\cite{HD2013}. Motivated by this, our work is focused in the special
choice of $f(R,G)$ that gives
\begin{equation}
\label{eq:lagrangiangb}
\mathcal{L}=\frac{1}{2}m_p^{2}\sqrt{-g}
\left( \alpha R^2+\beta G \log G \right)\,,
\end{equation}
for the Lagrangian with constants $\alpha, \beta$, where $m_p$ is
Planck's mass, $g$ is determinant of the metric tensor $g_{\mu\nu}$,
$R$ is Ricci scalar, and $G$ is Gauss-Bonnet invariant defined above.

The action is, as usual, defined as
\begin{equation}
\label{eq:action}
\mathcal{S}=\int{\mathcal{L} dx^4}\,.
\end{equation}

The paper is organised as follows: In Section \ref{background} we give
a comparison on how the Einstein tensor is written in GR, $f(R)$ and
$f(R,G)$ in general, then we present the background field equations
for the Lagrangian \eqref{eq:lagrangiangb}. In Section \ref{perturbed}
we present the linear order perturbed Einstein-like tensor that
describes the geometry of the universe, in longitudinal gauge for
scalar perturbations.
In Section \ref{conclusion} we conclude and give an outlook on future
work. The expression for the Einstein-like tensor, derived from a
general Lagrangian as a function $f(R,G)$ is given in Appendix
\ref{einstein}.

\textit{Notation.} The sign convention is $(-+++)$. Greek indices,
such as $\{\alpha, \beta, \dots, \mu, \nu, \dots\}$, run from 0 to
3. Latin indices, such as $\{ a,b,\dots, i,j,k,\dots \}$, run from 1
to 3, that is only over spatial dimensions. Throughout this cork we
use the units $c=\hbar=1$. We use prime to denote derivatives with
respect to conformal time, and we use a comma to denote partial
derivatives with respect to comoving spatial coordinates, i.e.,
\begin{equation}
X' \equiv \frac{\p X}{\p \eta}, \qquad X_{,i} \equiv \frac{\p X}{\p x^{i}}\,.
\end{equation}
For simplicity we work with a flat background spatial metric which is
compatible with current observations.

\section{Governing equations}
\label{background}

In standard GR, the Einstein-Hilbert action is given by integrating
the Lagrangian,
\begin{equation}
\label{eq:einstein-hilbert}
\mathcal{L}_{GR} = \frac{1}{2}m_{p}^{2} \sqrt{-g} R\,,
\end{equation}
and the field equations are obtained by varying the action with
respect to the metric tensor $g_{\mu \nu}$, in GR one has only the
Einstein tensor
\begin{equation}
G_{\mu \nu} = R_{\mu \nu} - \frac{1}{2} g_{\mu \nu} R\,,
\end{equation}
and one can then relate the geometry with the matter content of the
universe by making it equal to the energy momentum tensor and
obtaining the usual equations of motion (EOM), as
\begin{equation}
\label{eq:eomgen}
G_{\mu \nu} = \kappa^{2} T_{\mu \nu}\,,
\end{equation}
where $\kappa^{2} = 8 \pi G_{\rm N}$, $G_{\rm N}$ is Newton's
constant, and $T_{\mu \nu}$ the energy-momentum tensor. Similarly, in
the case of the most popular versions of modified gravity, one can
replace the Ricci tensor, $R$, in the Einstein-Hilbert action
\eqref{eq:einstein-hilbert} with a function $f(R)$ and get a modified
Einstein tensor,
\begin{equation}
\label{eq:fr}
\hat{G}_{\mu \nu} = R_{\mu \nu} \p_{R}f +g_{\mu \nu}\Bigg[
  \left(\nabla^{2}R\right) \p^{2}_{R}f +
  \left(\nabla_{\alpha}R\right)\left(\nabla^{\alpha}R\right)\p^{3}_{R}f
  -\frac{1}{2}f \Bigg] - \left(\nabla_{\mu}\nabla_{\nu}R\right)
\p^{2}_{R}f - \left(\nabla_{\mu}R\right) \left(\nabla_{\nu}R\right)
\p^{3}_{R}f\,,
\end{equation}
where we dropped the dependency on $R$ on $f(R)$ to keep the equation
more compact. We can see that \eqref{eq:fr} reduces to the usual
Einstein tensor if we set $f(R)=R$.\\

In principle one can use a function $f(R)$ as complex as one likes,
however, in this paper we concentrate on the modified gravity function
that includes the Gauss-Bonnet term $G$ defined in the previous
section in Eq.~\eqref{eq:gb}, and therefore we work with $f(R,G)$. The
full expression for the general Einstein-like tensor is rather lengthy
and is given in Appendix \ref{einstein}. Once more, if one drops the
dependency on $G$ and makes $f(R,G)=R$ only, one recovers the usual
Einstein tensor from GR. In this paper we use a particular function
$f(R,G)$ studied in Ref.~\cite{HD2013} given by
\begin{equation}
\label{eq:frg}
f(R,G)=\alpha R^2+\beta G \log G\,,
\end{equation}
where $\alpha$ and $\beta$ are arbitrary constants. This function
leads to a particular Einstein-like tensor
\begin{equation}
\label{eq:eom2}
\G_{\mu \nu} = 2 \alpha \left(R_{\mu \nu} R + g_{\mu \nu} \nabla^{2} R -\nabla_{\mu}\nabla_{\nu}R\right)-\frac{1}{2} \left( \alpha R^2+\beta G \log G \right)g_{\mu \nu}+\frac{\beta}{2}\left(1+\log G\right)\Big[\mathcal{C}\indices{^1_{\mu\nu}}\Big]+\frac{\beta}{2G}\Big[\mathcal{C}\indices{^2_{\mu\nu}}\Big]-\frac{\beta}{2G^{2}}\Big[\mathcal{C}\indices{^3_{\mu\nu}}\Big]\,,
\end{equation}
where the $\mathcal{C}\indices{^i_{\mu\nu}}$'s are the coefficients
corresponding to the derivatives $\p^{i}_{G}f$ from
Eq.~\eqref{eq:eom} \footnote{Note that $\p_{R}\p_{G}f(R,G)=\p_{G}\p_{R}f(R,G)=0$ with the choice for $f(R,G)$ made in Eq.~\eqref{eq:frg}.}, we will
be working with this description of the universe geometry.

By modifying the Lagrangian and adding the Gauss-Bonnet term, the governing
equations are also modified,
\begin{equation}
\label{mod_ein_equ}  
\G_{\mu \nu} = \kappa^{2}T_{\mu \nu}\,,
\end{equation}
analogous to Eq.~\eqref{eq:eomgen} where the right hand side stays the
same, since we are only modifying the way we describe the geometry of
the universe.\\

The perturbed Friedmann-Lemaitre-Robertson-Walker (FLRW) metric for a
universe with a flat background is given by \cite{KS,MFB,MM2008,
  malik}
\begin{equation}
\label{eq:pertmetric}
{\text{d}} s^2=a^{2}\big[-(1+2\phi){\text{d}}\eta^{2} + 2\left( B_{,i}-S_{i}\right){\text{d}}\eta {\text{d}} x^{i} + \big\{ (1-2\psi)\delta_{ij}+E_{,ij} +F_{(i,j)}+\frac{1}{2}h_{ij}\big\} {\text{d}} x^{i} {\text{d}} x^{j} \big]\,.
\end{equation}
where $a=a(\eta)$ is the scale factor, $\phi, B, \psi$ and $E$ are
\textit{scalar} metric perturbations, $S_{i}$ and $F_{i}$ are
\textit{vector} metric perturbations, and $h_{ij}$ is a
\textit{tensor} metric perturbation.
The reason for splitting the metric perturbation into these three
types is that the governing equations decouple at linear order, and
hence we can solve each perturbation type separately. At higher order
this is no longer true. In this paper we only study scalar
perturbations, postponing the discussion of vector and tensor
perturbations for future work \cite{FGM2}. \\

The general energy-momentum tensor for a fluid with density $\rho$, isotropic
pressure $P$ and fluid 4-velocity $u^{\mu}$, is defined as
\begin{equation}
T_{\mu \nu} = \left( \rho + P\right)u_{\mu}u_{\nu} + P g_{\mu \nu} +
\pi_{\mu \nu}\,.
\end{equation}
%
%
The fluid 4-velocity is subject to the constraint
\begin{equation}
u_{\mu}u^{\mu}=-1\,.
\end{equation}
The components of the 4-velocity $u_\mu$ up to linear order are then given
by
\bea
\label{defumu}
 u_0 &=& -a\left(1+\phi\right)\,, \nonumber \\
u_i &=& a\left(v_{,i}+B_{,i}
\right) \,,
\eea
where $v$ is the scalar velocity perturbation.

With the above definitions, the components of the stress energy tensor
in the background are given by
\begin{equation}
  {T^{0}}_{0} = -\rho_{0}\,, \qquad {T}^{0}_{i} = 0\,,
  \qquad {T^{i}}_{j} = {\delta^{i}}_{j}P_{0}\,,
\end{equation}
and at first order,
\begin{align}
\delta {T^{0}}_{0} &= -\delta \rho\,, \\
\delta {T^{0}}_{i} &= \left( \rho_{0}+P_{0}\right)\left( v_{,i}+B_{,i} \right)\,, \\
\delta {T^{i}}_{j} &= \delta P {\delta^{i}}_{j}+\Pi_{,~j}^{~i}\,,
\end{align}
where $\Pi$ is the scalar, anisotropic stress perturbation.

\subsection{Background}

In standard Einstein gravity, Eq.~\eqref{eq:eomgen}, 
the governing equations in component form can also be rewritten
as the Friedmann equations,
\begin{align}
\label{fried1}  
\H^{2} &= \frac{\kappa^{2}}{3}a^{2}\rho_{0}\,,\\
\label{fried2}
\H' &= -\frac{\kappa^{2}}{6}a^{2}\left( \rho_{0}+3 P_{0}\right)\,,
\end{align}
where $\H = a'/a$ is the Hubble parameter. Note that in this work we
use conformal time, related to cosmic time, $t$, by ${\text{d}} \tau = a {\text{d}}
\eta$.

Using the definition of the modified governing equations,
Equ.~\ref{mod_ein_equ} we can also find Friedmann-like equations for
$f(R,G)$ gravity. In the background these equations of motion are
given, for our choice of theory and hence $\G_{\mu \nu}$, by
\bea
\label{eq:g00bg}
&&\frac{18}{a^{2}}\Bigg[ \alpha\left(\H\H''-{\H'}^{2}-\H^{4}\right)+\frac{2}{3}\beta \left(66\H\H''-20{\H'}^{2}-44\H^{4}-31\H^{2}\H'\right.\\
&&\qquad\left.+\frac{1}{\H'}\left[35\H^{3}\H''-52\H^{6}-6{\H'}^{2}\right]-\frac{1}{\H^{2}}\left[4{\H'}^{3}+18{\H''}^{2}\right]+\frac{32}{\H}\H'\H''+\frac{16}{\H^{3}}{\H'}^{2}\H''-\frac{13}{\H^{4}}\H' {\H''}^{2}\right) \Bigg]=\kappa^2 T_{00}\,,\nonumber \\
\label{eq:gijbg}
&&\frac{6}{a^{2}} \delta_{ij}\Bigg[\alpha\left( \H\H''-\H'''-{\H'}^{2}+6\H^{2}\H'-\H^{4} \right) +\frac{2}{3}\beta\left(9\H^{2}\H'-2\H\H''+\frac{10}{3}{\H'}^{2}\right.\nonumber\\
&&\qquad\qquad\left.-\frac{2}{\H}\H'\H''+\frac{1}{\H'}\left[\H^{3}\H''-\H^{2}\H'''-\frac{4}{3}\H^{6}\right]+\frac{1}{{\H'}^{2}}\H^{2}{\H''}^{2}\right) \Bigg]=\kappa^2 T_{ij}\,.
\eea
The off-diagonal components of the field equations vanish in the
background. Also note that the logarithmic dependence $\log G$ cancels
out. 
Already at the background level in this theory the field equations,
(\ref{eq:g00bg}) and (\ref{eq:gijbg}), are more complicated than the
ones in the standard GR case, Eqs.~(\ref{fried1}) and
(\ref{fried2}). In particular the equations now contain time
derivatives up to fourth-order, instead of up to second-order in the
standard case.
The above equations have been previously derived in Ref.~\cite{nojiri:2018}, and our
results agree with the results found there.

\section{Perturbed governing equations}
\label{perturbed}

Due to the complexity of the governing equations in fourth-order
gravity, we derived the components of the Einstein-like tensor in
longitudinal gauge, instead of leaving the gauge unspecified. We can
easily reconstruct the tensor components for an arbitrary gauge, by
substituting in the definitions of the variables in longitudinal gauge
given in the following.

{Longitudinal gauge} is widely used in the literature, it has also
proven useful for calculations on small scales, since it gives
evolution equations closest to the Newtonian ones and it has also been
used in backreaction studies. The longitudinal gauge is completely
determined by the spatial gauge choice $\widetilde{E_{\ell}}=0$ and
$\widetilde{B_{\ell}}=0$. The remaining scalar metric perturbations,
$\phi$ and $\psi$, are given as
\begin{align}
\widetilde{\phi_{\ell}} &= \phi + \H\left(B+E'\right) + \left(B-E' \right)'\,,\\
\widetilde{\psi_{\ell}} &= \psi - \H\left( B-E' \right)\,,
\end{align}
note that $\widetilde{\phi_{\ell}}$ and $\widetilde{\psi_{\ell}}$ are
identical to the Bardeen potentials $\Phi$ and $\Psi$
\cite{Bardeen80}.

The fluid density perturbation, $\delta \rho$, and scalar velocity,
$v$, are given by
\begin{align}
  \widetilde{\delta \rho_{\ell}}
  &= \delta \rho + \rho'_{0}\left( B-E'\right)\,, \\
\tilde{v}_{\ell} &= v + E'\,.
\end{align}

These gauge-invariant quantities are simply a gauge-invariant
definition of the perturbations in the longitudinal gauge. Using these
definitions one can get the \textit{general} form of the field
equations. From here onwards we drop the tilde and the subscript so
the notation does not get cluttered.


\subsection{Scalars}

The scalar perturbations give the biggest contribution to the
components of the Einstein-like tensor, and this is given by
\begin{align}
\label{eq:g00n}
\delta\G_{00} &= -\frac{2}{a^{2}}\Bigg[ \alpha\Big\{18\left(\H^{4}+{\H'}^{2}-\H\H''\right)\phi-9\H\H'\phi'-9\H^{2}\phi''+9\left(4\H^{3}-\H\H'-\H''\right)\psi'\notag\\
&\qquad\qquad +9\left(2\H'-\H^{2}\right)\psi''-9\H\psi'''+12\left(\H^{2}+\H'\right)\nabla^{2}\phi+15\H\nabla^{2}\psi'+3\nabla^{2}\psi''+\nabla^{2}\nabla^{2}\left(\phi-2\psi\right) \Big\}\notag\\
&\qquad\quad +\beta\Big\{ \left( 24\H^{2}\left[26\H^{8}+19\H^{6}\H'+20\H^{4}{\H'}^{2}+7\H^{2}{\H'}^{3}+2{\H'}^{4}\right]-6\H\left[73\H^{6}+126\H^{4}\H'\right.\right.\notag\\
&\qquad\qquad\quad\left.\left.+64\H^{2}{\H'}^{2}+32{\H'}^{3}\right]\H''+12\left[6\H^{4}+18\H^{2}\H'+13{\H'}^{2}\right]{\H''}^{2} \right)\frac{\phi}{\H^{4}\H'}\notag\\
&\qquad\qquad+\left(6\H\left[ 13{\H'}^{2}-6\H^{4} \right]{\H''}^{2} -3\H\left[104\H^{10}+4\H^{8}\H'+139\H^{6}{\H'}^{2}+322\H^{4}{\H'}^{3}+168\H^{2}{\H'}^{4}\right.\right.\notag\\
&\qquad\qquad\quad\left.\left.+96{\H'}^{5}\right]+6\left[ 35\H^{8}+\H^{6}\H'+4\H^{4}{\H'}^{2}+76\H^{2}{\H'}^{3}+78{\H'}^{4} \right]\H'' \right)\frac{\phi'}{\H^{4}{\H'}^{2}}\notag\\
&\qquad\qquad-\left( 71\H^{7}+130\H^{5}\H'+64\H^{3}{\H'}^{2}+32\H{\H'}^{3}-24\H^{4}\H''-72\H^{2}\H'\H'' -52{\H}^{2}\H'' \right)\frac{3\phi''}{\H^{3}\H'}\notag\\
&\qquad\qquad-\left( 104\H^{12}-620\H^{10}\H'-189\H^{8}{\H'}^{2}+174\H^{6}{\H'}^{3}+176\H^{4}{\H'}^{4}+112\H^{2}{\H'}^{5} -70\H^{9}\H''\right.\notag\\
&\qquad\qquad\quad +213\H^{7}\H'\H''+118\H^{5}{\H'}^{2}\H'' -216\H^{3}{\H'}^{3}\H''-252\H{\H'}^{4}\H''+12\H^{6}{\H''}^{2}+46\H^{2}{\H'}^{2}{\H''}^{2}\notag\\
&\qquad\qquad\quad\left.+104{\H'}^{3}{\H''}^{2}\right)\frac{3\psi'}{\H^{5}{\H'}^{2}}\notag\\
&\qquad\qquad-\left( 104\H^{10}+75\H^{8}\H'+56\H^{6}{\H'}^{2}-4\H^{4}{\H'}^{3}+8\H^{2}{\H'}^{4}-70\H^{7}\H''-26\H^{5}\H'\H''-8\H^{3}{\H'}^{2}\H''\right.\notag\\
&\qquad\qquad\quad\left. +12\H{\H'}^{3}\H''+12\H^{4}{\H''}^{2}-26{\H'}^{2}{\H''}^{2} \right)\frac{3\psi''}{\H{\H'}^{2}}\notag\\
&\qquad\qquad-\left( 71\H^{7}+130\H^{5}\H'+64\H^{3}{\H'}^{2}+32\H{\H'}^{3}-24\H^{4}\H''-72\H^{2}\H'\H'' -52{\H}^{2}\H'' \right)\frac{3\psi'''}{\H^{4}\H'}\notag\\
&\qquad\qquad+\left( 4\H^{2}\left[ 18{\H'}^{4}+44\H^{2}{\H'}^{3}+77\H^{4}{\H'}^{2}+51\H^{6}\H'-26^{8} \right] +2\left[ 13{\H'}^{2}-6\H^{4} \right]{\H''}^{2} \right.\notag\\
&\qquad\qquad\quad\left. +2\H\left[ 35\H^{6}-34\H^{4}\H' -109\H^{2}{\H'}^{2}-68{\H'}^{3}\right]\H'' \right)\frac{\nabla^{2}\phi}{\H^{4}{\H'}^{2}}\notag\\
&\qquad\qquad-\left( 70\H^{7}+130\H^{5}\H' +66\H^{3}{\H'}^{2}+32\H{\H'}^{3}-24\H^{4}\H''-73\H^{2}\H'\H''-52{\H'}^{2}\H''\right)\frac{\nabla^{2}\phi'}{\H^{4}\H'}\notag\\
&\qquad\qquad-\left( 104\H^{10}+42\H^{8}\H' +2\H^{6}{\H'}^{2}-20\H^{4}{\H'}^{3}-8\H^{2}{\H'}^{4}-36\H^{7}\H''+\H^{5}\H'\H''+32\H^{3}{\H'}^{2}\H''\right.\notag\\
&\qquad\qquad\quad\left.+32\H{\H'}^{3}\H''-18\H^{2}\H'{\H''}^{2}-26{\H'}^{2}{\H''}^{2} \right)\frac{4\nabla^{2}\psi}{\H^{6}\H'}\notag\\
&\qquad\qquad+\left( 209\H^{7}+236\H^{5}\H'+100\H^{3}{\H'}^{2}+48\H{\H'}^{3}-70\H^{4}\H''-152\H^{2}\H'\H''-72{\H'}^{2}\H'' \right)\nabla^{2}\frac{\psi'}{\H^{4}\H'}\notag\\
&\qquad\qquad+\left( \frac{\H^{2}}{\H'} \right)\nabla^{2}\psi''+\left( \frac{\H^{2}}{3\H'} \right)\nabla^{2}\nabla^{2}\phi-\left( \frac{2}{3} \right)\nabla^{2}\nabla^{2}\psi  \Big\} \Bigg]\,,
\end{align}
\begin{align}
\label{eq:g0in}
\delta \G_{0i} = \delta\G_{i0} &= \frac{2}{a^{2}}\Bigg[\alpha\Big\{3\left(3\H''+2\H\H'-4\H^{3}\right)\phi-3\left( \H^{2}-3\H' \right)\phi'+3\H\phi''+3\left(7\H'-5\H^{2}\right)\psi'+3\psi'''\notag\\
&\qquad\qquad\quad-3\H\nabla^{2}\left(\phi-2\psi\right)+\nabla^{2}\left(\phi'-2\psi'\right)\Big\}+\beta\Big\{ \left(60{\H'}^{3}\H''+171\H^{4}\H'\H'' +172\H^{4}\H'\H'' \right.\notag\\
&\qquad\qquad\quad\left. +44\H^{6}\H''-72\H{\H'}^{4} -134\H^{3}{\H'}^{3}-220\H^{5}{\H'}^{2}-282\H^{7}\H'-192\H^{9}\right)\frac{\nabla^{2}\phi}{6\H^{4}{\H'}^{2}} \notag\\
&\qquad\qquad+\left( \right. 96\H^{9}+156\H^{7}\H' +84\H^{5}{\H'}^{2}+83\H^{3}{\H'}^{3}+24\H{\H'}^{4}-24\H^{6}\H''-82\H^{4}\H'\H'' \notag\\
&\qquad\qquad\quad \left. -90\H^{2}{\H'}^{2}\H''-24{\H'}^{3}\H'' \right)\frac{2\nabla^{2}\psi}{3\H^{4}{\H'}^{2}}\notag\\
&\qquad\qquad-\left( 1880\H^{9}\H'+2156\H^{7}{\H'}^{2} +1710\H^{5}{\H'}^{3}+992\H^{3}{\H'}^{4}+432\H{\H'}^{5}+576\H^{8}\H''+178\H\H'\H''\right. \notag\\
&\qquad\qquad\quad -1231\H^{4}{\H'}^{2}\H''-984\H^{2}{\H'}^{3}\H'' -360{\H'}^{4}\H''-144\H^{5}{\H''}^{2}-444\H^{3}\H'{\H''}^{2} \notag\\
&\qquad\qquad\quad\left.-288\H{\H'}^{2}{\H''}^{2}\right)\frac{\phi}{6\H^{4}{\H'}^{2}}\notag\\
&\qquad\qquad-\left( 288\H^{9}+419\H^{7}\H'+323\H^{5}{\H'}^{2} +176\H^{3}{\H'}^{3}+108\H{\H'}^{4}-66\H^{6}\H''-258\H^{4}\H'\H''\right.\notag\\
&\qquad\qquad\quad\left. -246\H^{2}{\H'}^{2}\H''-90{\H'}^{3}\H'' \right)\frac{\phi'}{3\H^{3}{\H'}^{2}}\notag\\
&\qquad\qquad-\left( 576\H^{10}+1566\H\H'+1126\H^{6}{\H'}^{2}+734\H^{4}{\H'}^{3}+504\H^{2}{\H'}^{4}+444\H^{7}\H''-56\H^{5}\H'\H'' \right. \notag\\
&\qquad\qquad\quad \left. -691\H^{3}{\H'}^{2}\H'' -180\H{\H'}^{3}\H'' -144\H^{4}{\H''}^{2}-444\H^{2}\H'{\H''}^{2}-288{\H'}^{2}{\H''}^{2} \right)\frac{\psi'}{6\H^{4}{\H'}^{2}}\notag\\
&\qquad\qquad-\left( 144\H^{9}+208\H^{7}\H'+166\H^{5}{\H'}^{2}+88\H^{3}{\H'}^{3}+54\H{\H'}^{4}-33\H^{6}\H''-129\H^{4}\H'\H'' \right. \notag\\
&\qquad\qquad\quad \left. -123\H^{2}{\H'}^{2}\H''-45{\H'}^{3}\H'' \right)\frac{2\psi''}{3\H^{4}{\H'}^{2}}\notag\\
&\qquad\qquad+\left( \frac{\H^{2}}{3\H'} \right)\nabla^{2}\phi'-\left(\frac{2}{3}\right)\nabla^{2}\psi'+\left( \frac{\H^{3}}{\H'} \right)\phi''+\left( \frac{\H^{2}}{\H'} \right)\psi'''  \Big\} \Bigg]_{,i}\,, 
\end{align}
\begin{align}
\label{eq:gijn}
\delta \G_{ij} &= \frac{2}{a^{2}}\Bigg[ \alpha\Big\{\delta_{ij}\bigg[ 12\left(\H^{4}-6\H^{2}\H'+{\H'}^{2}-\H\H''+\H'''\right)\phi-3\left( 6\H^{3}+\H\H'-6\H'' \right)\phi' -3\left(\H^{2}-4\H'\right)\phi''\notag\\
&\qquad\qquad +3\H\phi'''+6\left(\H^{4}-6\H^{2}\H'+{\H'}^{2}-\H\H''+\H'''\right)\psi -3\left( 2\H^{3}+13\H\H'-5\H'' \right)\psi'+3\psi''''\notag \\ 
&\qquad\qquad\quad-3\left(7\H^{2}-6\H'\right)\psi'' -2\left( \H^{2}+2\H' \right)\nabla^{2}\phi-3\H\nabla^{2}\left(2\phi'+\psi'\right)+\nabla^{2}\left(\phi''-5\psi''\right)\notag\\
&\qquad\qquad\quad-2\left(\H^{2}-\H'\right)\nabla^{2}\psi -\nabla^{2}\nabla^{2}\left(\phi-2\psi\right)\bigg]+\bigg[ 3\H\left(\phi'+3\psi'\right) + 6\left( \H^{2}+\H' \right)\psi \bigg]_{,ij}\notag\\
&\qquad\qquad\quad+\bigg[ 3\psi''+\nabla^{2}\nabla^{2}\left( \phi-2\psi \right) \bigg]_{,ij} \Big\} \notag\\
&\qquad\qquad+ \beta\Big\{ \delta_{ij}\bigg[ \left( 8\H^{3}\H'\H'' -16 \H^{4}{\H'}^{2}-4{\H'}^{4}-4\H{\H'}^{2}\H''+\H^{2}\left[13{\H'}^{3}-{\H''}^{2}\right] \right)\frac{8\phi}{{\H'}^{2}}\notag\\
&\qquad\qquad-\left( 8\H^{8}\H' +48\H^{6} {\H'}^{2}+64\H^{2}{\H'}^{4} -18{\H'}^{5}-6\H^{5}\H'\H''+36\H^{3}{\H'}^{2}\H''\right.\notag\\
&\qquad\qquad\quad \left. -3\H^{4}\left[ 47{\H'}^{3}+4{\H''}^{2}-2\H'\H''' \right] \right)\frac{\phi'}{3\H{\H'}^{3}}+\left( 7\H^{4}\H'+2\H^{2}{\H'}^{2}+2{\H'}^{3}-4\H^{3}\H'' \right)\frac{\phi''}{{\H'}^{2}}\notag\\
&\qquad\qquad+\left( 4\H^{7}\H'-48\H^{5}{\H'}^{2}-22\H{\H'}^{4}+21\H^{4}\H'\H''-6\H^{2}{\H'}^{2}\H''+6{\H'}^{3}\H''\right.\notag\\
&\qquad\qquad\quad\left. +3\H^{3}\left[ 4{\H'}^{3}-2{\H''}^{2}+\H'\H'' \right] \right)\frac{2\psi}{3\H{\H'}^{2}}\notag\\
&\qquad\qquad-\left( 8\H^{9}\H' +32\H^{7}{\H'}^{2}-14\H{\H'}^{5}-6\H^{6}\H'\H'' -81\H^{4}{\H'}^{2}\H'' + 42\H^{2}{\H'}^{3}\H'' +18{\H'}^{4}\H''\right.\notag\\
&\qquad\qquad\quad \left. -\H^{3}\left[ 92{\H'}^{4}-12\H'{\H''}^{2} \right]+3\H^{5}\left[ 49{\H'}^{3}-4{\H''}^{2}+2\H'\H'' \right] \right)\frac{\psi'}{3\H^{2}{\H'}^{3}}\notag\\
&\qquad\qquad-\left( 8\H^{6}\H'+27\H^{4}{\H'}^{2}+40{\H'}^{4}+6\H^{3}\H'\H''+18\H{\H'}^{2}\H''-6\H^{2}\left[ 14{\H'}^{3}+2{\H''}^{2}-\H'\H''' \right] \right)\frac{\psi''}{3{\H'}^{3}}\notag\\
&\qquad\qquad+\left( 4\H^{4}\H'-\H^{2}{\H'}^{2}+{\H'}^{3}-2\H^{3}\H'' \right)\frac{2\psi'''}{\H{\H'}^{2}}\notag\\
&\qquad\qquad-\left( 16\H^{7}\H' +30\H^{5}{\H'}^{2}+131\H{\H'}^{4}-12\H^{4}\H'\H''+6\H^{2}{\H'}^{2}\H''-6{\H'}^{3}\H''\right. \notag\\
&\qquad\qquad\quad\left.-3\H^{3}\left[7{\H'}^{3}+{\H''}^{2}-4\H'\H'''\right] \right)\frac{\nabla^{2}\phi}{18\H{\H'}^{3}}+\left( 7\H^{4}\H' -7\H^{2}{\H'}^{2} +{\H'}^{3} -4\H^{3}\H'' \right)\frac{\nabla^{2}\phi'}{3\H{\H'}^{2}}\notag\\
&\qquad\qquad+\left( 2\H^{7}\H'+453\H^{5}{\H'}^{2}-68\H{\H'}^{4} -234\H^{4}\H'\H''+78\H^{2}{\H'}^{2}\H'' +48{\H'}^{3}\H''\right. \notag \\
&\qquad\qquad\quad\left. -3\H^{3}\left[ 69{\H'}^{3}-12{\H''}^{2}+2\H'\H''' \right] \right)\frac{\nabla^{2}\psi}{18\H^{{\H'}^{2}}}+\left( 4\H^{4}-21\H^{2}\H'+6{\H'}^{2}+4\H\H'' \right)\frac{\nabla^{2}\psi'}{3\H\H'}\notag\\ 
&\qquad\qquad+\left( \frac{\H^{2}}{\H'} \right)\psi''''+\left( \frac{\H^{3}}{\H'} \right)\phi'''+\left( \frac{\H^{2}}{3\H'} \right)\nabla^{2}\phi''-\left( \frac{5}{3} \right)\nabla^{2}\psi''-\left( \frac{1}{3} \right)\nabla^{2}\nabla^{2}\phi+\left( \frac{2\H'}{3\H^{2}} \right)\nabla^{2}\nabla^{2}\psi \bigg] \notag\\
&\qquad\qquad+\bigg[\left( 11\H^{3}\H'-10\H^{5}+23\H{\H'}^{2}-2\H^{2}\H''-6\H'\H'' \right)\frac{\phi}{2\H\H'}\notag\\
&\qquad\qquad +\left( 2\H^{6}\H'-79\H^{4}{\H'}^{2}-36{\H'}^{4}+42\H^{3}\H'\H''-6\H{\H'}^{2}\H''+\H^{2}\left[ 53{\H'}^{3}-12{\H''}^{2}+6\H'\H'' \right] \right)\frac{\psi}{6\H^{2}{\H'}^{2}}\notag\\
&\qquad\qquad+\left( \H \right)\phi' -\left( 4\H^{5}+\H^{3}\H' -26\H{\H'}^{2}+6\H'\H''\right)\frac{\psi'}{3\H^{2}\H'}+\psi'' +\left( \frac{1}{3} \right)\nabla^{2}\phi-\left( \frac{2\H'}{\H^{2}} \right)\nabla^{2}\psi \bigg]_{,ij} \Big\}\Bigg]\,.
\end{align}

The trace of the spatial part of the Einstein-like tensor is given by
\begin{align}
\label{eq:tracegn}
\delta {\G^{k}}_{k} &= \frac{2}{a^{2}}\Bigg[ \alpha\Big\{ 36\left(\H^{4}-6\H^{2}\H'+{\H'}^{2}-\H\H''+\H'''\right)\phi-9\left( 6\H^{3}+\H\H' -6\H'' \right)\phi'-9\left( \H^{2}-4\H' \right)\phi''\notag\\
&\qquad\qquad+9 \H \phi''' +18\left( \H^{4}-6\H^{2}\H'+{\H'}^{2}-\H\H''+\H''' \right)\psi-9\left( 2\H^{3}+13\H\H'-5\H'' \right)\psi'\notag\\
&\qquad\qquad-9\left( 7\H^{2}-6\H' \right)\psi'' +9\psi''''-6\left( \H^{2}+2\H' \right)\nabla^{2}\phi-15\H\nabla^{2}\phi'+3\nabla^{2}\phi''+12 \H' \nabla^{2}\psi \notag\\
&\qquad\qquad-12\nabla^{2}\psi''-2\nabla^{2}\nabla^{2}\phi+4\nabla^{2}\nabla^{2}\psi \Big\}\notag\\
&\qquad\qquad+\beta\Big\{\left( 8\H^{3}\H'\H''-16\H^{4}{\H'}^{2}-4{\H'}^{4}-4\H{\H'}^{2}\H''+\H^{2}\left[ 13{\H'}^{3}-{\H''}^{2} \right] \right)\frac{24\phi}{{\H'}^{2}}\notag\\ 
&\qquad\qquad-\left( 8\H^{8}\H'+48\H^{6}{\H'}^{2}+64\H^{2}{\H'}^{4}-18{\H'}^{5}-6\H^{5}\H'\H''+36\H^{3}{\H'}^{2}\H'' \right.\notag\\
&\qquad\qquad\quad\left.-3\H^{4}\left[ 47{\H'}^{3}+4{\H''}^{2} -2\H'\H''' \right]\right)\frac{\phi'}{\H{\H'}^{3}}\notag\\
&\qquad\qquad+\left( 7\H\H'+2\H^{2}{\H'}^{2}+2{\H'}^{3}-4\H^{3}\H'' \right)\frac{3\phi''}{{\H'}^{2}}\notag\\
&\qquad\qquad+\left( 4\H^{7}\H'+48\H^{5}{\H'}^{2}-22\H{\H'}^{4}+21\H^{4}\H'\H'' -6\H^{2}{\H'}^{2}+6{\H'}^{3}\H''\right.\notag\\
&\qquad\qquad\quad\left. +3\H^{3}\left[ 4{\H'}^{3}-2{\H''}^{2}+\H'\H''' \right] \right)\frac{2\psi}{\H{\H'}^{2}}\notag\\
&\qquad\qquad-\left( 8\H^{6}\H'+27\H^{4}{\H'}^{2}+40{\H'}^{4}+6\H^{3}\H'\H''+18\H{\H'}^{2}\H''\right.\notag\\
&\qquad\qquad\quad\left.-6\H^{2}\left[ 14{\H'}^{3}+2{\H''}^{2}-\H'\H''' \right] \right)\frac{\psi''}{{\H'}^{3}}+\left( 4\H^{4}\H' -\H^{2}{\H'}^{2}+{\H'}^{3}-2\H^{3}\H'' \right)\frac{6\psi'''}{\H{\H'}^{2}}\notag\\
&\qquad\qquad-\left( 8\H^{7}\H' +30\H^{5}{\H'}^{2}-27\H^{3}{\H'}^{3}+31\H{H'}^{4}+6\H'\left[ {\H'}^{2}+\H^{2}\H'-\H^{4} \right]\right.\notag\\
&\qquad\qquad\quad\left. -12\H^{3}{\H''}^{2}+6\H^{3}\H'\H'''  \right)\frac{\nabla^{2}\phi}{3\H{\H'}^{3}}+\left( \H'\left[ 7\H^{4}-6\H^{2}\H'+2{H'}^{2} \right]-4\H^{3}\H'' \right)\nabla^{2}\phi'\notag\\
&\qquad\qquad+\left( 2\H^{7}\H' +187\H^{5}{\H'}^{2}-52\H{\H'}^{4}-96\H^{4}\H'\H''+36\H^{2}{\H'}^{2}\H'' +24{\H'}^{3}\H''\right.\notag\\
&\qquad\qquad\quad \left. +\H^{3}\left[ 12{\H''}^{2}-77{\H'}^{3} \right]  \right)\frac{\nabla^{2}\psi}{3\H^{3}{\H'}^{2}}\notag\\
&\qquad\qquad+\left( 2\H\left[ 2\H^{4}-16\H^{2}\H' +11{\H'}^{2} \right]+\left[ 6\H^{2}-3\H' \right]\H'' \right)\frac{2\nabla^{2}\psi'}{3\H^{2}\H'}\notag\\
&\qquad\qquad+\left(\frac{\H^{3}}{\H'}\right)\phi'''+\left(\frac{3\H^{2}}{\H'}\right)\psi''''+\left( \frac{\H^{2}}{\H'} \right)\nabla^{2}\phi''-4\nabla^{2}\psi''-\left(\frac{2}{3}\right)\nabla^{2}\nabla^{2}\phi+\left( \frac{4\H'}{3\H^{2}} \right)\nabla^{2}\nabla^{2}\psi\Big\} \Bigg]\,.
\end{align}

To construct the traceless Einstein-like tensor we need to subtract
the trace from the spatial part given by Eq.~\eqref{eq:tracegn}. In
order to do this, we apply two spatial derivatives to $\delta \G_{ij}$
(see e.g.~Ref.~\cite{malik}). We then use the linearity of the equation
to simplify it even more, and we are left with single Laplacian
operators instead of double Laplacians, simplifying calculations,
%
\begin{align}
\label{eq:iintg}
\iint \left(\p^{i}\p^{j}\left[\delta \G_{ij}\right]\right) &= \frac{2}{a^{2}}\Bigg[  \alpha\Big\{ \left(\H^{4}-6\H^{2}\H'{\H}^{2}-\H\H''+\H'''\right)\phi-3\left( 6\H^{3}+\H\H'-6\H'' \right)\phi' -3\left( \H^{2}-4\H' \right)\phi'' \notag\\ 
&\qquad\qquad+3\H\phi'''+6\left( \H^{4}-6\H^{2}\H'+{\H}^{2}-\H\H'' \right) \psi-3\left( 2\H^{3}+13\H\H'-5\H'' \right)\psi'\notag\\ 
&\qquad\qquad-3\left( 7\H^{2}-6\H' \right)\psi'' +3\psi''''-2\left(\H^{2}+2\H'\right)\nabla^{2}\left(\phi-2\psi\right)-3\H\nabla^{2}\left(\phi' -2\psi' \right)\notag\\ 
&\qquad\qquad+\nabla^{2}\left(\phi''-2\psi''\right) \Big\} \notag\\ 
&\qquad\qquad+ \beta\Big\{\left(8\H^{3}\H'\H''-16\H^{4}{\H'}^{2}-4{\H'}^{4}-4\H{\H'}^{2}\H''+\H^{2}\left[ 13{\H'}^{3}-{\H''}^{2} \right]\right)\frac{8\phi}{{\H'}^{2}}\notag\\ 
&\qquad\qquad-\left( 8\H^{8}\H'+48\H^{6}{\H'}^{2}+64\H^{2}{\H'}^{4}-18{\H'}^{5}-6\H^{5}\H'\H''+36\H^{3}{\H'}^{2}\H''\right.\notag\\
&\qquad\qquad\quad\left. -3\H^{4}\left[ 47{\H'}^{3}+4{\H''}^{2}-2\H'\H''' \right] \right)\frac{\phi'}{3\H {\H'}^{3}}\notag\\ 
&\qquad\qquad+\left( 7\H^{4}\H' +2\H^{2}{\H'}^{2}+2{\H'}^{3} - 4\H^{3}\H'' \right)\frac{\phi''}{{\H'}^{2}}\notag\\ 
&\qquad\qquad+\left( 4\H^{7}\H' - 48\H^{5}{\H'}^{2} - 22\H{\H'}^{4} + 21\H^{4}\H'\H''-6{\H'}^{3}\H''\right.\notag \\
&\qquad\qquad\quad\left. +3\H^{3}\left[ 4{\H'}^{3}-2{\H''}^{2}+\H'\H''' \right]  \right)\frac{2\psi}{3\H{\H'}^{2}}\notag\\ 
&\qquad\qquad-\left( 8\H^{9}\H' + 32\H^{7}{\H'}^{2} -14\H{\H'}^{5} -14\H{\H'}^{5} -6\H^{6}\H'\H''-81\H^{4}{\H'}^{2}\H'' +42\H^{2}{\H'}^{2}\H''\right.\notag\\
&\qquad\qquad\quad\left. +\H^{3}\left[ 12\H'{\H''}^{2} -92{\H'}^{4} \right] +3\H^{5}\left[ 49{\H'}^{3}-4{\H''}^{2} +2\H'\H''' \right] \right)\frac{\psi'}{3\H^{2}{\H'}^{3}}\notag\\ 
&\qquad\qquad-\left( 8\H^{6}\H' +27\H^{4}{\H'}^{2}+40{\H'}^{4}+6\H^{3}\H'\H'' +18\H {\H'}^{2} \H'' \right.\notag\\
&\qquad\qquad\quad \left. -6\H^{2}\left[ 14{\H'}^{3} +2{\H''}^{2} -\H'\H''' \right] \right)\frac{\psi''}{3{\H'}^{3}}\notag\\ 
&\qquad\qquad+\left( 4\H^{4}\H' -\H^{2}{\H'}^{2}+{\H'}^{3}-2\H^{3}\H'' \right)\frac{2\psi'''}{\H{\H'}^{2}}\notag\\ 
&\qquad\qquad-\left( 4\H^{7}\H'+30\H^{5}{\H'}^{2}-19\H{\H'}^{4}-3\H^{4}\H'\H''+6\H^{2}{\H'}^{2}\H''+12{\H'}^{3}\H'' \right.\notag\\
&\qquad\qquad\quad\left. +\H^{3}\left[ 3\H'\H'''-6{\H''}^{2}-30{\H'}^{3} \right] \right)\frac{2\nabla^{2}\phi}{9\H{\H'}^{3}}\notag\\ 
&\qquad\qquad+\left( 7\H^{4}\H'-4\H^{2}{\H'}^{2}+2{\H'}^{3}-4\H^{3}\H'' \right)\frac{ \nabla^{2}\phi'}{3\H{\H'}^{2}}\notag\\ 
&\qquad\qquad+\left( 2^{7}+54^{5}\H' -44\H{\H'}^{3} -27\H^{4}\H'' +15\H^{2}\H'\H'' + 12{\H'}^{2}\H''\right.\notag\\
&\qquad\qquad\quad\left. +3\H^{3}\left[ \H'''-4{\H'}^{2} \right] \right)\frac{\nabla^{2}\psi}{81\H^{6}{\H'}^{4}}\notag\\ 
&\qquad\qquad-\left(11\H^{3}\H' -16\H{\H'}^{2} - 2\H^{2}\H'' + 3\H' \H'' \right) \frac{\nabla^{2}\psi'}{27\H^{5}{\H'}^{4}} \notag\\ 
&\qquad\qquad+\left( \frac{\H^{3}}{\H'} \right)\phi'''+\left( \frac{\H^{2}}{{\H'}^{2}} \right)\psi''''+\left( \frac{\H^{2}}{3\H'}\right)\nabla^{2}\phi''-\frac{\nabla^{2}\psi''}{27\H^{3}{\H'}^{3}}\Big\} \Bigg]\,.
\end{align}

Subtracting \eqref{eq:iintg} and \eqref{eq:tracegn} from
Eq.~\eqref{eq:gijn} one gets the trace free Einstein-like tensor.




\section{Conclusion and Future Work}
\label{conclusion}

In this paper, we have provided a derivation of the governing
equations for a fourth order $f(R,G)$ theory. Throughout we assumed a
flat FRLW universe. We rederived the governing equations in the
background, and, using cosmological perturbation theory to linear
order, derived the governing equations for the scalar perturbations.
Since the equations are rather complex, we only present them in
longitudinal gauge. However, we also provide the definitions of the
gauge-invariant variables in this gauge, both for the metric and the
matter perturbations. One can therefore easily rewrite the governing
equations for any other gauge.


Here we only studied scalar perturbations, leaving the discussion of
vector and tensor perturbations for future work \cite{FGM2}. The
recent detection of gravitational waves and present and future
gravitational wave observatories like LIGO \cite{Abbott:2007kv},
Virgo\cite{Abbott:2009kk}, KAGRA \cite{KAGRA}, and LISA
\cite{AmaroSeoane:2012je} makes calculating the tensor perturbation
evolution equations an exciting future project.

We also need to find solutions to the governing equations, since only
then can we calculate observational signatures that can be compared to
the observational data. The equations presented in this work are
rather complex and hence difficult to solve. In order to make solving
the equations less time consuming, we have also provided the
Mathematica sheets in Github\cite{githubGB}.

\begin{acknowledgements}
The authors are grateful to Pedro Carrilho and Timothy Clifton for
useful discussions and comments. KAM is supported in part by the STFC
under grants ST/M001202/ and ST/P000592/1.~JLF acknowledges support of
studentship funded by Queen Mary University of London and CONACYT
grant No.~603085.~UAG acknowledges support of the IRSIP fellowship of
Higher Education Commission (HEC) Pakistan. The authors acknowledge
the use of the tensor algebra package \texttt{xAct} \cite{xact} and
its subpackage \texttt{xPand} \cite{xpand}.
\end{acknowledgements}



\appendix
\section{Einstein-like tensor}
\label{einstein}

The choice of a function of the Ricci scalar is arbitrary, so we keep the function only as $f(R,G)$ without any assumption, we do take into account the fact that the Gauss-Bonnet term $G$, given in Eq.~\eqref{eq:gb}, also has a dependency on the metric, so it has to be varied with respect of the metric itself, from where the equation becomes rather lengthy, this general equation, reduces to the known cases of GR and $f(R)$ when we choose $f(R)=R$ and we drop the dependance of $G$ respectively.

In general, the Einstein-like tensor, without choosing any particular $f(R,G)$, is given by
\begin{align}
\label{eq:eom}
\tilde{G}_{\mu \nu} &= g_{\mu \nu} \Big\{ \p_{R}f\Big[0\Big]+\p^{2}_{R}f\Big[ \nabla^{2}R \Big]+\p^{3}_{R}f\Big[ \left( \nabla_{\alpha}R\right)\left( \nabla^{\alpha}R \right)\Big] +\p_{G}f\Big[2 \nabla^{2}R-4\nabla_{\alpha}\nabla_{\beta}R^{\alpha \beta }\Big] -\frac{1}{2}f \notag \\
&\quad+\p^{2}_{G}f\Big[ 4R^{2}\left( \nabla^{2}R \right)-32R^{\beta\sigma}\left( \nabla_{\alpha}R_{\beta \sigma} \right)\left( \nabla^{\alpha}R \right)+12R\left(\nabla_{\alpha}R\right)\left(\nabla^{\alpha}R\right) + 8R^{\beta\sigma\lambda\rho}\left(\nabla_{\alpha}R_{\beta\sigma\lambda\rho}\right)\left(\nabla^{\alpha}R\right) \notag \\
&\qquad \quad -16R \left(\nabla^{\alpha}R\right)\left(\nabla_{\beta}R\indices{_\alpha^\beta}\right) + 32 R^{\alpha \beta}\left(\nabla_{\alpha}R^{\sigma \lambda}\right)\left(\nabla_{\beta}R_{\sigma\lambda}\right)-16R^{\sigma\lambda\rho\tau}\left(\nabla_{\alpha}R^{\alpha\beta}\right)\left(\nabla_{\beta}R_{\sigma\lambda\rho\tau}\right) \notag \\
&\qquad \quad -8R^{\alpha \beta} \left( \nabla_{\alpha}R^{\sigma\lambda\rho\tau} \right)\left(\nabla_{\beta}R_{\sigma\lambda\rho\tau}\right) -8 R^{\alpha\beta}R \left(\nabla_{\alpha}\nabla_{\beta}R\right)-8R^{\alpha \beta}\left(R^{\sigma\lambda\rho\tau}\right)\left(\nabla_{\alpha}\nabla_{\beta}R_{\sigma\lambda\rho\tau}\right) -8R_{\alpha\beta}\left(\nabla^{\alpha}R\right)\left(\nabla^{\beta}R\right) \notag \\
&\qquad \quad -16R^{\alpha\beta} R \left( \nabla^{2}R_{\alpha \beta} \right)-16R \left(\nabla_{\sigma}R_{\alpha\beta}\right)\left(\nabla^{\sigma}R^{\alpha\beta}  \right) +64R^{\alpha\beta}\left(\nabla^{\sigma}R_{\alpha\beta}\right)\left(\nabla_{\lambda}R\indices{_\sigma^\lambda}\right)+32R^{\alpha\beta}R^{\sigma\lambda}\left(\nabla_{\lambda}\nabla_{\sigma}R_{\alpha\beta}\right)\notag \\
&\qquad \quad +4R^{\alpha\beta\sigma\lambda}R\left(\nabla^{2}R_{\alpha\beta\sigma\lambda}\right) +4 R \left(\nabla_{\rho}R_{\alpha\beta\sigma\lambda}\right)\left(\nabla^{\rho}R^{\alpha\beta\sigma\lambda}\right)\Big] \notag \\
&\quad+\p^{3}_{G}f\Big[ 8R^{3}\left(\nabla_{\alpha}R\right)\left(\nabla^{\alpha}R\right)-64R^{\sigma\beta}R^{2}\left(\nabla_{\alpha}R_{\beta\sigma}\right)\left(\nabla^{\alpha}R\right)+16R^{2}R^{\beta\sigma\lambda\rho}\left(\nabla_{\alpha}R_{\beta\sigma\lambda\rho}\right)\left(\nabla^{\alpha}R\right) \notag \\
&\qquad \quad +128 R\indices{_\alpha^\beta}R^{\sigma\lambda}R\left(\nabla^{\alpha}R\right)\left(\nabla_{\beta}R_{\sigma\lambda}\right)-32R\indices{_\alpha^\beta}R^{\sigma\lambda\rho\tau}R\left(\nabla^{\alpha}R\right)\left(\nabla_{\beta}R_{\sigma\lambda\rho\tau}\right) \notag \\
&\qquad \quad -16R^{\alpha\beta}R^{\sigma\lambda\rho\tau}R^{\xi\chi\omega\kappa}\left(\nabla_{\alpha}R_{\sigma\lambda\rho\tau}\right)\left(\nabla_{\beta}R_{\xi\chi\omega\kappa}\right)-16R_{\alpha\beta}R^{2}\left(\nabla^{\alpha}R\right)\left(\nabla^{\beta}R\right) \notag\\
&\qquad \quad -64R^{\alpha\beta}R^{\lambda\rho\xi\chi}R\left(\nabla_{\sigma}R_{\lambda\rho\xi\chi}\right)\left(\nabla^{\sigma}R_{\alpha\beta}\right)-256R^{\alpha\beta}R^{\sigma\lambda}R^{\rho\tau}\left(\nabla_{\sigma}R_{\alpha\beta}\right)\left(\nabla_{\lambda}R_{\rho\tau}\right)\notag \\
&\qquad \quad +128R^{\alpha \beta}R^{\sigma\lambda}R^{\rho\tau\xi\chi}\left(\nabla_{\sigma}R_{\alpha \beta}\right)\left(\nabla_{\lambda}R_{\rho\tau\xi\chi}\right) +128 R^{\alpha \beta}R^{\sigma\lambda}R\left(\nabla_{\rho}R_{\sigma\lambda}\right)\left(\nabla^{\rho}R_{\alpha\beta}\right)\notag\\
&\qquad \quad+8R^{\alpha\beta\sigma\lambda}R^{\rho\tau\xi\chi}R\left(\nabla_{\kappa}R_{\rho\tau\xi\chi}\right)\left(\nabla^{\kappa}R_{\alpha\beta\sigma\lambda}\right) \Big] \notag \\
&\quad+\p_{R}\p_{G}f\Big[4 R\left( \nabla^{2}R \right)+6 \left( \nabla_{\alpha}R\right)\left( \nabla^{\alpha}R \right)-8\left( \nabla^{\alpha}R\right)\left(\nabla_{\beta}R\indices{_\alpha^\beta} \right)-4R^{\alpha \beta}\left( \nabla_{\alpha}\nabla_{\beta}R\right)-8R^{\alpha \beta}\left( \nabla^{2}R_{\alpha \beta}\right) \notag \\
&\qquad \quad-8\left( \nabla_{\sigma}R_{\alpha \beta} \right)\left( \nabla^{\sigma} R^{\alpha \beta}\right)+2R^{\alpha \beta \lambda \rho} \left( \nabla^{2}R_{\alpha \beta \lambda \rho}\right)+2\left( \nabla_{\sigma}R_{\alpha \beta \lambda \rho}\right)\left( \nabla^{\sigma}R^{\alpha \beta \lambda \rho} \right)\Big] \notag \\
&\quad+\p^{2}_{R}\p_{G}f\Big[ 6R\left( \nabla_{\alpha}R\right)\left( \nabla^{\alpha}R \right) -16R^{\beta \sigma}\left( \nabla_{\alpha}R_{\beta \sigma}\right)\left( \nabla^{\alpha} R\right) + 4R^{\beta \sigma \lambda \rho}\left( \nabla_{\alpha}R_{\beta \sigma\lambda \rho} \right)\left( \nabla^{\alpha}R\right) - 4R_{\alpha \beta}\left( \nabla^{\alpha}R\right)\left( \nabla^{\beta}R\right)\Big] \notag \\
&\quad+\p_{R}\p^{2}_{G}f\Big[ 12 R^{2}\left( \nabla_{\alpha}R\right)\left( \nabla^{\alpha}R\right) -64R^{\beta \sigma}R \left( \nabla_{\alpha}R_{\beta \sigma} \right)\left( \nabla^{\alpha}R \right)+16 R^{\beta \sigma \lambda \rho}R\left( \nabla_{\alpha}R_{\beta \sigma \lambda \rho}\right)\left( \nabla^{\alpha}R \right)\notag\\ 
&\qquad \quad +64R\indices{_\alpha^\beta}R^{\sigma \lambda} \left( \nabla^{\alpha}R \right)\left( \nabla_{\beta}R_{\sigma \lambda} \right) -16 R\indices{_\alpha^\beta}R^{\sigma\lambda\rho\tau}\left( \nabla^{\alpha}R\right)\left( \nabla_{\beta}R_{\sigma\lambda\rho\tau}\right)-16R_{\alpha \beta}R \left( \nabla^{\alpha}R\right)\left( \nabla^{\beta}R\right) \notag \\
&\qquad \quad -32R^{\alpha \beta} R^{\lambda\rho\tau\xi}\left( \nabla_{\sigma}R_{\lambda\rho\tau\xi}\right)\left( \nabla^{\sigma}R_{\alpha \beta} \right)+64R^{\alpha \beta} R^{\sigma \lambda}\left( \nabla_{\rho}R_{\sigma\lambda}\right)\left( \nabla^{\rho}R_{\alpha \beta}\right)+4R^{\alpha \beta \sigma \lambda}R^{\rho \tau\xi\kappa}\left( \nabla_{\chi}R_{\rho\tau\xi\kappa} \right)\left( \nabla^{\chi}R_{\alpha\beta\sigma\lambda} \right)\Big]\notag \\
&\qquad \Big\} \notag 
\end{align}
\begin{align}
&\qquad + \p_{R}f\Big[ R_{\mu \nu} \Big]-\p^{2}_{R}f\Big[ \left(\nabla_{\mu}\nabla_{\nu}R\right) \Big]-\p^{3}_{R}f\Big[ \left(\nabla_{\mu}R\right)\left(\nabla_{\nu}R\right) \Big]\notag\\ 
&\quad\quad+\p_{G}f\Big[ 2R_{\mu\nu}R-8R\indices{_\mu^\alpha}R_{\nu\alpha}+2R\indices{_\mu^\alpha^\beta^\sigma}R_{\nu\alpha\beta\sigma} -4\left(\nabla^{2}R_{\mu\nu}\right)+2\left(\nabla_{\alpha}\nabla_{\beta}R\indices{_\mu^\alpha_\nu^\beta}\right)+4\left(\nabla_{\alpha}\nabla_{\mu}R\indices{_\nu^\alpha}\right) +4\left(\nabla_{\alpha}\nabla_{\nu}R\indices{_\mu^\alpha}\right)\notag \\
&\qquad \quad +2\left(\nabla_{\beta}\nabla_{\alpha}R\indices{_\mu^\alpha_\nu^\beta}\right)-2\left(\nabla_{\nu}\nabla_{\mu}R\right)\Big]\notag \\
&\quad\quad+\p_{R}\p_{G}f\Big[ 4\left(\nabla^{\alpha}R\right)\left(\nabla_{\beta}R\indices{_\mu_\alpha_\nu^\beta}\right)-4R_{\mu\nu}\left(\nabla^{2}R\right) -8\left(\nabla_{\alpha}R_{\mu\nu}\right)\left(\nabla^{\alpha}R\right)+4\left(\nabla^{\alpha}R\right)\left(\nabla_{\beta}R\indices{_\mu^\beta_\nu_\alpha}\right)\notag \\
&\qquad\quad +4R_{\mu\alpha\nu\beta}\left(\nabla^{\beta}\nabla^{\alpha}R\right) +4\left(\nabla^{\alpha}R\right)\left(\nabla_{\mu}R_{\nu\alpha}\right)+4\left(\nabla_{\alpha}R\indices{_\nu^\alpha}\right)\left(\nabla_{\mu}R\right)+4R\indices{_\nu^\alpha}\left(\nabla_{\mu}\nabla_{\alpha}R\right)+4R^{\alpha\beta}\left(\nabla_{\mu}\nabla_{\nu}R_{\alpha\beta}\right)\notag\\
&\qquad \quad - R^{\alpha\beta\sigma\lambda}\left(\nabla_{\mu}\nabla_{\nu}R_{\alpha\beta\sigma\lambda}\right)+8\left(\nabla_{\mu}R^{\alpha\beta}\right)\left(\nabla_{\nu}R_{\alpha\beta}\right)+4\left(\nabla^{\alpha}R\right)\left(\nabla_{\nu}R_{\mu\alpha}\right)+4\left(\nabla_{\alpha}R\indices{_\mu^\alpha}\right)\left(\nabla_{\nu}R\right)\notag\\
&\qquad \quad -6\left(\nabla_{\mu}R\right)\left(\nabla_{\nu}R\right)-2\left(\nabla_{\mu}R^{\alpha\beta\sigma\lambda}\right)\left(\nabla_{\nu}R_{\alpha\beta\sigma\lambda}\right)+4R\indices{_\mu^\alpha}\left(\nabla_{\nu}\nabla_{\alpha}R\right)+4R^{\alpha\beta}\left(\nabla_{\mu}\nabla_{\nu}R_{\alpha\beta}\right)\notag\\
&\qquad\quad -4R\left(\nabla_{\mu}\nabla_{\mu}R\right) -R^{\alpha\beta\sigma\lambda}\left(\nabla_{\mu}\nabla_{\nu}R_{\alpha\beta\sigma\lambda}\right)\Big]\notag\\ 
&\quad\quad+\p^{2}_{R}\p_{G}f\Big[4 R_{\mu\alpha\nu\beta}\left(\nabla^{\alpha}R\right)\left(\nabla^{\beta}R\right)-4R_{\mu\nu}\left(\nabla_{\alpha}R\right)\left(\nabla^{\alpha}R\right)+4R_{\nu\alpha}\left(\nabla^{\alpha}R\right)\left(\nabla_{\mu}R\right)+8R^{\alpha\beta}\left(\nabla_{\mu}R\right)\left(\nabla_{\nu}R_{\alpha\beta}\right) \notag\\
&\qquad \quad +4R_{\mu\alpha}\left(\nabla^{\alpha}R\right)\left(\nabla_{\nu}R\right) +8R^{\alpha \beta}\left(\nabla_{\mu}R_{\alpha\beta}\right)\left(\nabla_{\nu}R\right)-6R\left(\nabla_{\mu}R\right)\left(\nabla_{\nu}R\right)-2R^{\alpha\beta\sigma\lambda}\left(\nabla_{\mu}R_{\alpha\beta\sigma\lambda}\right)\left(\nabla_{\nu}R\right)\notag\\
&\qquad\quad -2R^{\alpha\beta\sigma\lambda}\left(\nabla_{\mu}R\right)\left(\nabla_{\nu}R_{\alpha\beta\sigma\lambda}\right)\Big]\notag
&\quad\quad+\p^{2}_{G}f\Big[ 4 R^{\sigma\lambda\rho\tau}R\indices{_\mu^\alpha_\nu^\beta}\left(\nabla_{\alpha}\nabla_{\beta}R_{\sigma\lambda\rho\tau}\right)-8R_{\mu\nu}R\left(\nabla^{2}R\right) -32R^{\beta\sigma}R\indices{_\nu^\alpha}\left(\nabla_{\alpha}\nabla_{\mu}R_{\beta\sigma}\right)+8R\indices{_\nu^\alpha}R^{\beta\sigma\lambda\rho}\left(\nabla_{\alpha}\nabla_{\mu}R_{\beta\sigma\lambda\rho}\right)\notag \\
&\qquad \quad -32R^{\beta\sigma}R\indices{_\mu^\alpha}\left(\nabla_{\alpha}\nabla_{\nu}R_{\beta\sigma}\right)+8R\indices{_\mu^\alpha}R^{\beta\sigma\lambda\rho}\left(\nabla_{\alpha}\nabla_{\nu}R_{\beta\sigma\lambda\rho}\right)-16R^{\beta\sigma\lambda\rho}\left(\nabla_{\alpha}R_{\beta\sigma\lambda\rho}\right)\left(\nabla^{\alpha}R_{\mu\nu}\right)-16 R \left(\nabla_{\alpha}R_{\mu\nu}\right)\left(\nabla^{\alpha}R\right)\notag \\
&\qquad \quad -8R_{\mu\nu}\left(\nabla_{\alpha}R\right)\left(\nabla^{\alpha}R\right)+8R\indices{_\mu^\alpha_\nu^\beta}\left(\nabla_{\alpha}R^{\sigma\lambda\rho\tau}\right)\left(\nabla_{\beta}R_{\sigma\lambda\rho\tau}\right)+8R \left(\nabla^{\alpha}R\right)\left(\nabla_{\beta}R\indices{_{\mu\alpha\nu}^\beta}\right)+8R \left(\nabla^{\alpha}R\right)\left(\nabla_{\beta}R\indices{_\mu^\beta_{\nu\alpha}}\right)\notag \\
&\qquad \quad +4R^{\sigma\lambda\rho\tau}R\indices{_\mu^\alpha_\nu^\beta}\left(\nabla_{\beta}\nabla_{\alpha}R_{\sigma\lambda\rho\tau}\right)+8R_{\mu\alpha\nu\beta}\left(\nabla^{\alpha}R\right)\left(\nabla^{\beta}R\right)+8R_{\mu\alpha\nu\beta}\left(\nabla^{\beta}\nabla^{\alpha}R\right)+32R^{\alpha\beta}R_{\mu\nu}\left(\nabla^{2}R_{\alpha\beta}\right)\notag\\
&\qquad \quad +32R_{\mu\nu}\left(\nabla_{\sigma}R_{\alpha\beta}\right)\left(\nabla^{\sigma}R^{\alpha\beta}\right)+64R^{\alpha\beta}\left(\nabla_{\sigma}R_{\alpha\beta}\right)\left(\nabla^{\sigma}R_{\mu\nu}\right)-32R^{\alpha\beta}\left(\nabla^{\sigma}R_{\alpha\beta}\right)\left(\nabla_{\lambda}R\indices{_{\mu\sigma\nu}^\lambda}\right)\notag \\ 
&\qquad \quad -32R^{\alpha\beta}\left(\nabla^{\sigma}R_{\alpha\beta}\right)\left(\nabla_{\lambda}R\indices{_\mu^\lambda_\nu_\sigma}\right)-32R_{\mu\sigma\nu\lambda}\left(\nabla^{\sigma}R^{\alpha\beta}\right)\left(\nabla^{\lambda}R_{\alpha\beta}\right)-16R^{\alpha\beta}R_{\mu\sigma\nu\lambda}\left(\nabla^{\lambda}\nabla^{\sigma}R_{\alpha\beta}\right) \notag \\
&\qquad \quad -16R^{\alpha\beta}R_{\mu\lambda\nu\sigma}\left(\nabla^{\lambda}\nabla^{\sigma}R_{\alpha\beta}\right)-32R^{\alpha\beta}\left(\nabla_{\sigma}R\indices{_\nu^\sigma}\right)\left(\nabla_{\mu}R_{\alpha\beta}\right)-32R\indices{_\nu^\alpha}\left(\nabla_{\alpha}R_{\beta\sigma}\right)\left(\nabla_{\mu}R^{\beta\sigma}\right)+8R\left(\nabla^{\alpha}R\right)\left(\nabla_{\mu}R_{\nu\alpha}\right)\notag \\ 
&\qquad \quad +8R^{\beta\sigma\lambda\rho}\left(\nabla_{\alpha}R_{\beta\sigma\lambda\rho}\right)\left(\nabla_{\mu}R\indices{_\nu^\alpha}\right)-32R^{\alpha\beta}\left(\nabla_{\sigma}R_{\alpha\beta}\right)\left(\nabla_{\mu}R\indices{_\nu^\sigma}\right)+8R\left(\nabla_{\alpha}R\indices{_\nu^\alpha}\right)\left(\nabla_{\mu}R\right)+8R_{\nu\alpha}\left(\nabla^{\alpha}R\right)\left(\nabla_{\mu}R\right)\notag \\
&\qquad \quad +8R^{\beta\sigma\lambda\rho}\left(\nabla_{\alpha}R\indices{_\nu^\alpha}\right)\left(\nabla_{\mu}R_{\beta\sigma\lambda\rho}\right)+8R\indices{_\nu^\alpha}\left(\nabla_{\alpha}R_{\beta\sigma\lambda\rho}\right)\left(\nabla_{\mu}R^{\beta\sigma\lambda\rho}\right)+8R\indices{_\nu^\alpha}R\left(\nabla_{\mu}\nabla_{\alpha}R\right)+8R^{\alpha\beta}R\left(\nabla_{\mu}\nabla_{\nu}R_{\alpha\beta}\right)\notag \\
&\qquad \quad -2R^{\alpha\beta\sigma\lambda}R\left(\nabla_{\mu}\nabla_{\nu}R_{\alpha\beta\sigma\lambda}\right) -32R^{\alpha\beta}\left(\nabla_{\sigma}R\indices{_\mu^\sigma}\right)\left(\nabla_{\nu}R_{\alpha\beta}\right)+16R\left(\nabla_{\mu}R^{\alpha\beta}\right)\left(\nabla_{\nu}R_{\alpha\beta}\right)+16R^{\alpha\beta}\left(\nabla_{\mu}R\right)\left(\nabla_{\nu}R_{\alpha\beta}\right)\notag \\
&\qquad \quad -32R\indices{_\mu^\alpha}\left(\nabla_{\alpha}R_{\beta\sigma}\right)\left(\nabla_{\nu}R^{\beta\sigma}\right)+8R\left(\nabla^{\alpha}R\right)\left(\nabla_{\nu}R_{\mu\alpha}\right)+8R^{\beta\sigma\lambda\rho}\left(\nabla_{\alpha}R_{\beta\sigma\lambda\rho}\right)\left(\nabla_{\nu}R\indices{_\mu^\alpha}\right)-32R^{\alpha\beta}\left(\nabla_{\sigma}R_{\alpha\beta}\right)\left(\nabla_{\nu}R\indices{_\mu^\sigma}\right)\notag \\
&\qquad\quad +8R \left(\nabla_{\alpha}R\indices{_\mu^\alpha}\right)\left(\nabla_{\nu}R\right)+8R_{\mu\alpha}\left(\nabla^{\alpha}R\right)\left(\nabla_{\nu}R\right)+16R^{\alpha\beta}\left(\nabla_{\mu}R_{\alpha\beta}\right)\left(\nabla_{\nu}R\right)-12R\left(\nabla_{\mu}R\right)\left(\nabla_{\nu}R\right)\notag\\
&\qquad \quad-4R^{\alpha\beta\sigma\lambda}\left(\nabla_{\mu}R_{\alpha\beta\sigma\lambda}\right)\left(\nabla_{\nu}R\right) -4R^{\alpha\beta\sigma\lambda}\left(\nabla_{\mu}R\right)\left(\nabla_{\nu}R_{\alpha\beta\sigma\lambda}\right)-4R\left(\nabla_{\mu}R^{\alpha\beta\sigma\lambda}\right)\left(\nabla_{\nu}R_{\alpha\beta\sigma\lambda}\right)\notag\\
&\qquad\quad +8R^{\beta\sigma\lambda\rho}\left(\nabla_{\alpha}R\indices{_\mu^\alpha}\right)\left(\nabla_{\nu}R_{\beta\sigma\lambda\rho}\right)+8R\indices{_\mu^\alpha}\left(\nabla_{\alpha}R_{\beta\sigma\lambda\rho}\right)\left(\nabla_{\nu}R^{\beta\sigma\lambda\rho}\right)+8R\indices{_\mu^\alpha}R\left(\nabla_{\nu}\nabla_{\alpha}R\right)+8R^{\alpha\beta}R\left(\nabla_{\nu}\nabla_{\mu}R_{\alpha\beta}\right)\notag \\
&\qquad \quad -4R^{2}\left(\nabla_{\nu}\nabla_{\mu}R\right)-2R^{\alpha\beta\sigma\lambda}R\left(\nabla_{\mu}\nabla_{\nu}R_{\alpha\beta\sigma\lambda}\right)-8R_{\mu\nu}R^{\alpha\beta\sigma\lambda}\left(\nabla^{2}R_{\alpha\beta\sigma\lambda}\right)-8R_{\mu\nu}\left(\nabla_{\rho}R_{\alpha\beta\sigma\lambda}\right)\left(\nabla^{\rho}R^{\alpha\beta\sigma\lambda}\right)\notag \\
&\qquad \quad +8R^{\alpha\beta\sigma\lambda}\left(\nabla_{\rho}R\indices{_\mu^\rho_\nu^\tau}\right)\left(\nabla_{\tau}R_{\alpha\beta\sigma\lambda}\right)+8R^{\alpha\beta\sigma\lambda}\left(\nabla_{\rho}R_{\alpha\beta\sigma\lambda}\right)\left(\nabla^{\rho}R\indices{_\mu^\rho_\nu^\tau}\right)  \Big]\notag\\ 
&\quad\quad+\p^{3}_{G}f\Big[ 128R^{\beta\sigma}R_{\mu\nu}R\left(\nabla_{\alpha}R_{\beta\sigma}\right)\left(\nabla^{\alpha}R\right) -16R_{\mu\nu}R^{2}\left(\nabla_{\alpha}R\right)\left(\nabla^{\alpha}R\right)-32R_{\mu\nu}R^{\beta\sigma\lambda\rho}R\left(\nabla_{\alpha}R_{\beta\sigma\lambda\rho}\right)\left(\nabla^{\alpha}R\right)\notag\\
&\qquad\quad +16R^{\sigma\lambda\rho\tau}R\indices{_{\mu\alpha\nu}^\beta}R\left(\nabla^{\alpha}R\right)\left(\nabla_{\beta}R_{\sigma\lambda\rho\tau}\right)+16R^{\sigma\lambda\rho\tau}R\indices{_\mu^\beta_{\nu\alpha}}R\left(\nabla^{\alpha}R\right)\left(\nabla_{\beta}R_{\sigma\lambda\rho\tau}\right)\notag \\
&\qquad \quad +16R^{\sigma\lambda\rho\tau}R\indices{_\mu^\alpha_\nu^\beta}R^{\xi\chi\kappa\omega}\left(\nabla_{\alpha}R_{\sigma\lambda\rho\lambda}\right)\left(\nabla_{\beta}R_{\xi\chi\kappa\omega}\right)+16R^{2}R_{\mu\alpha\nu\beta}\left(\nabla^{\alpha}R\right)\left(\nabla^{\beta}R\right)+128R^{\alpha\beta}R_{\mu\nu}R^{\lambda\rho\tau\xi}\left(\nabla_{\sigma}R_{\lambda\rho\tau\xi}\right)\left(\nabla^{\sigma}R_{\alpha\beta}\right)\notag\\
&\qquad\quad -64R^{\alpha\beta}R\indices{_{\mu\sigma\nu}^\lambda}R^{\rho\tau\xi\chi}\left(\nabla^{\sigma}R_{\alpha\beta}\right)\left(\nabla_{\lambda}R_{\rho\tau\xi\chi}\right)-64R^{\alpha\beta}R\indices{_\mu^\lambda_{\nu\sigma}}R^{\rho\tau\xi\chi}\left(\nabla^{\sigma}R_{\alpha\beta}\right)\left(\nabla_{\lambda}R_{\rho\tau\xi\chi}\right)\notag \\
&\qquad \quad -64R^{\beta\sigma} R_{\mu\alpha\nu\lambda}R\left(\nabla^{\alpha}R\right)\left(\nabla^{\lambda}R_{\beta\sigma}\right)-64R^{\beta\sigma}R_{\mu\lambda\nu\alpha}R\left(\nabla^{\alpha}R\right)\left(\nabla^{\lambda}R_{\beta\sigma}\right)+256R^{\beta\sigma}R^{\lambda\rho}R\indices{_\nu^\alpha}\left(\nabla_{\alpha}R_{\lambda\rho}\right)\left(\nabla_{\mu}R_{\beta\sigma}\right)\notag \\
&\qquad\quad-64R^{\beta\sigma}R\indices{_\nu^\alpha}R^{\lambda\rho\tau\xi}\left(\nabla_{\alpha}R_{\lambda\rho\tau\xi}\right)\left(\nabla_{\mu}R_{\beta\sigma}\right)-64R^{\beta\sigma}R_{\nu\alpha}R\left(\nabla^{\alpha}R\right)\left(\nabla_{\mu}R_{\beta\sigma}\right)-64R^{\beta\sigma}R\indices{_\nu^\alpha}R\left(\nabla_{\alpha}R_{\beta\sigma}\right)\left(\nabla_{\mu}R\right)\notag \\
&\qquad\quad+16R\indices{_\nu^\alpha}R^{\beta\sigma\lambda\rho}R\left(\nabla_{\alpha}R_{\beta\sigma\lambda\rho}\right)\left(\nabla_{\mu}R\right)+16R_{\nu\alpha}R^{2}\left(\nabla^{\alpha}R\right)\left(\nabla_{\mu}R\right)+16R\indices{_\nu^\alpha}R^{\beta\sigma\lambda\rho}R^{\tau\xi\chi\kappa}\left(\nabla_{\alpha}R_{\tau\xi\chi\kappa}\right)\left(\nabla_{\mu}R_{\beta\sigma\lambda\rho}\right)\notag\\
&\qquad \quad +16R_{\nu\alpha}R^{\beta\sigma\lambda\rho}R\left(\nabla^{\alpha}R\right)\left(\nabla_{\mu}R_{\beta\sigma\lambda\rho}\right)-64R^{\beta\sigma}R\indices{_\nu^\alpha}R^{\lambda\rho\tau\xi}\left(\nabla_{\alpha}R_{\beta\sigma}\right)\left(\nabla_{\mu}R_{\lambda\rho\tau\xi}\right)+32R^{\alpha\beta}R^{2}\left(\nabla_{\mu}R\right)\left(\nabla_{\nu}R_{\alpha\beta}\right)\notag\\
&\qquad \quad+32R^{\alpha\beta}R^{\sigma\lambda\rho\tau}R\left(\nabla_{\mu}R_{\sigma\lambda\rho\tau}\right)\left(\nabla_{\nu}R_{\alpha\beta}\right)+256R^{\beta\sigma}R^{\lambda\rho}R\indices{_\mu^\alpha}\left(\nabla_{\alpha}R_{\lambda\rho}\right)\left(\nabla_{\nu}R_{\beta\sigma}\right)-64R^{\beta\sigma}R\indices{_\mu^\alpha}R^{\lambda\rho\tau\xi}\left(\nabla_{\alpha}R_{\lambda\rho\tau\xi}\right)\left(\nabla_{\nu}R_{\beta\sigma}\right)\notag\\
&\qquad\quad-64R^{\beta\sigma}R_{\mu\alpha}R\left(\nabla^{\alpha}R\right)\left(\nabla_{\nu}R_{\beta\sigma}\right)-128R^{\alpha\beta}R^{\sigma\lambda}R\left(\nabla_{\mu}R_{\alpha\beta}\right)\left(\nabla_{\nu}R_{\sigma\lambda}\right)-64 R^{\beta\sigma}R\indices{_\mu^\alpha}R\left(\nabla_{\alpha}R_{\beta\sigma}\right)\left(\nabla_{\nu}R\right)\notag\\
&\qquad\quad+16R\indices{_\mu^\alpha}R^{\beta\sigma\lambda\rho}R\left(\nabla_{\alpha}R_{\beta\sigma\lambda\rho}\right)\left(\nabla_{\nu}R\right)+16R_{\mu\alpha}R^{2}\left(\nabla^{\alpha}R\right)\left(\nabla_{\nu}R\right)+32R^{\alpha\beta}R^{2}\left(\nabla_{\mu}R_{\alpha\beta}\right)\left(\nabla_{\nu}R\right)\notag\\
&\qquad\quad-8R^{3}\left(\nabla_{\mu}R\right)\left(\nabla_{\nu}R\right)-8R^{2}R^{\alpha\beta\sigma\lambda}\left(\nabla_{\mu}R_{\alpha\beta\sigma\lambda}\right)\left(\nabla_{\nu}R\right)-8R^{2}R^{\alpha\beta\sigma\lambda}\left(\nabla_{\mu}R\right)\left(\nabla_{\nu}R_{\alpha\beta\sigma\lambda}\right)\notag\\
&\qquad\quad+16R\indices{_\mu^\alpha}R^{\beta\sigma\lambda\rho}R^{\tau\xi\chi\kappa}\left(\nabla_{\alpha}R_{\tau\xi\chi\kappa}\right)\left(\nabla_{\nu}R_{\beta\sigma\lambda\rho}\right) +16R_{\mu\alpha}R^{\beta\sigma\lambda\rho}R\left(\nabla^{\alpha}R\right)\left(\nabla_{\nu}R_{\beta\sigma\lambda\rho}\right)\notag\\
&\qquad\quad+32R^{\alpha\beta}R^{\sigma\lambda\rho\tau}R\left(\nabla_{\mu}R_{\alpha\beta}\right)\left(\nabla_{\nu}R_{\sigma\lambda\rho\tau}\right)-64R^{\beta\sigma}R\indices{_\mu^\alpha}R^{\lambda\rho\tau\xi}\left(\nabla_{\alpha}R_{\beta\sigma}\right)\left(\nabla_{\nu}R_{\lambda\rho\tau\xi}\right)\notag\\
&\qquad\quad-8R^{\alpha\beta\sigma\lambda}R^{\rho\xi\chi\kappa}R\left(\nabla_{\mu}R_{\alpha\beta\sigma\lambda}\right)\left(\nabla_{\nu}R_{\rho\xi\chi\kappa}\right)-256R^{\alpha\beta}R^{\sigma\lambda}R_{\mu\nu}\left(\nabla_{\rho}R_{\sigma\lambda}\right)\left(\nabla^{\rho}R_{\alpha\beta}\right)\notag\\
&\qquad\quad +256R^{\alpha\beta}R^{\sigma\lambda}R_{\mu\rho\nu\tau}\left(\nabla^{\rho}R_{\alpha\beta}\right)\left(\nabla^{\tau}R_{\sigma\lambda}\right)-16R_{\mu\nu}R^{\alpha\beta\sigma\lambda}R^{\rho\tau\xi\chi}\left(\nabla_{\kappa}R_{\rho\tau\xi\chi}\right)\left(\nabla^{\kappa}R_{\alpha\beta\sigma\lambda}\right) \Big]\notag
\end{align}
\begin{align}
&\quad\quad+\p_{R}\p^{2}_{G}f\Big[ 64R^{\beta\sigma}R_{\mu\nu}\left(\nabla_{\alpha}R_{\beta\sigma}\right)\left(\nabla^{\alpha}R\right)-16R_{\mu\nu}\left(\nabla_{\alpha}R\right)\left(\nabla^{\alpha}R\right)-16R_{\mu\nu}R^{\beta\sigma\lambda\rho}\left(\nabla_{\alpha}R_{\beta\sigma\lambda\rho}\right)\left(\nabla^{\alpha}R\right)\notag\\
&\qquad\quad +8R^{\sigma\lambda\rho\tau}R\indices{_\mu_\alpha_\nu^\beta}\left(\nabla^{\alpha}R\right)\left(\nabla_{\beta}R_{\sigma\lambda\rho\tau}\right)+8R^{\sigma\lambda\rho\tau}R\indices{_\mu^\beta_{\nu\alpha}}\left(\nabla^{\alpha}R\right)\left(\nabla_{\beta}R_{\sigma\lambda\rho\tau}\right)+16R_{\mu\alpha\nu\beta}R\left(\nabla^{\alpha}R\right)\left(\nabla^{\beta}R\right)\notag \\
&\qquad\quad -32R^{\beta\sigma}R_{\mu\alpha\nu\lambda}\left(\nabla^{\alpha}R\right)\left(\nabla^{\lambda}R_{\beta\sigma}\right)-32R^{\beta\sigma}R_{\mu\lambda\nu\alpha}\left(\nabla^{\alpha}R\right)\left(\nabla^{\lambda}R_{\beta\sigma}\right)-32R^{\beta\sigma}R_{\nu\alpha}\left(\nabla^{\alpha}R\right)\left(\nabla_{\mu}R_{\beta\sigma}\right)\notag\\
&\qquad\quad -32R^{\beta\sigma}R\indices{_\nu^\alpha}\left(\nabla_{\alpha}R_{\beta\sigma}\right)\left(\nabla_{\mu}R\right)+8R\indices{_\nu^\alpha}R^{\beta\sigma\lambda\rho}\left(\nabla_{\alpha}R_{\beta\sigma\lambda\rho}\right)\left(\nabla_{\mu}R\right)+16R_{\nu\alpha}R\left(\nabla^{\alpha}R\right)\left(\nabla_{\mu}R\right)\notag \\
&\qquad\quad +8R_{\nu\alpha}R^{\beta\sigma\lambda\rho}\left(\nabla^{\alpha}R\right)\left(\nabla_{\mu}R_{\beta\sigma\lambda\rho}\right)+32R^{\alpha\beta}R\left(\nabla_{\mu}R\right)\left(\nabla_{\nu}R_{\alpha\beta}\right)+16R^{\alpha\beta}R^{\sigma\lambda\rho\tau}\left(\nabla_{\mu}R_{\sigma\lambda\rho\tau}\right)\left(\nabla_{\nu}R_{\alpha\beta}\right)\notag \\
&\qquad \quad -32R^{\beta\sigma}R_{\mu\alpha}\left(\nabla^{\alpha}R\right)\left(\nabla_{\nu}R_{\beta\sigma}\right)-64R^{\alpha\beta}R^{\sigma\lambda}\left(\nabla_{\mu}R_{\alpha\beta}\right)\left(\nabla_{\nu}R_{\sigma\lambda}\right)-32R^{\beta\sigma}R\indices{_\mu^\alpha}\left(\nabla_{\alpha}R_{\beta\sigma}\right)\left(\nabla_{\nu}R\right)\notag \\
&\qquad\quad +8R\indices{_\mu^\alpha}R^{\beta\sigma\lambda\rho}\left(\nabla_{\alpha}R_{\beta\sigma\lambda\rho}\right)\left(\nabla_{\nu}R\right) +16R_{\mu\alpha}R\left(\nabla^{\alpha}R\right)\left(\nabla_{\nu}R\right)+32R^{\alpha\beta}R\left(\nabla_{\mu}R_{\alpha\beta}\right)\left(\nabla_{\nu}R\right)-12R^{2}\left(\nabla_{\mu}R\right)\left(\nabla_{\nu}R\right)\notag \\
&\qquad \quad -8R^{\alpha\beta\sigma\lambda}R\left(\nabla_{\mu}R_{\alpha\beta\sigma\lambda}\right)\left(\nabla_{\nu}R\right)-8R^{\alpha\beta\sigma\lambda}R\left(\nabla_{\mu}R_{\alpha\beta}\right)\left(\nabla_{\nu}R\right)-8R^{\alpha\beta\sigma\lambda}R\left(\nabla_{\mu}R\right)\left(\nabla_{\nu}R_{\alpha\beta\sigma\lambda}\right)\notag\\
&\qquad\quad +8R_{\mu\alpha}R^{\beta\sigma\lambda\rho}\left(\nabla^{\alpha}R\right)\left(\nabla_{\nu}R_{\beta\sigma\lambda\rho}\right)+16R^{\alpha\beta}R^{\sigma\lambda\rho\tau}\left(\nabla_{\mu}R_{\alpha\beta}\right)\left(\nabla_{\nu}R_{\sigma\lambda\rho\tau}\right)-4R^{\alpha\beta\sigma\lambda}R^{\rho\tau\xi\chi}\left(\nabla_{\mu}R_{\alpha\beta\sigma\lambda}\right)\left(\nabla_{\nu}R_{\rho\tau\xi\chi}\right)\Big]
\end{align}
where $\p_{R}$ and $\p_{G}$ are the derivatives with respect to the Ricci scalar and the Gauss-Bonnet term, respectively, i.e. $\p_{R} = \p/\p R$ and $\p_{G} = \p/\p G$. From here also one can identify the $\mathcal{C}\indices{^i_{\mu\nu}}$ present in Eq.~\eqref{eq:eom2}.


\begin{thebibliography}{}


\bibitem{b}
S. Nojiri, S. D. Odintsov,  
Int. J. Geom. Meth. Mod. Phys. 4, 115 (2007)]
 [arXiv:hepth/0601213].

\bibitem{c}
S. Nojiri, S. D. Odintsov,
 arxiv:1011.0544 [gr-qc]
 
\bibitem{LDS2007}
 L. Amendola, D. Polarski and S. Tsujikawa,
 Phys. Rev. Lett. 98, 131302 (2007).

\bibitem{DES}
	J.~Frieman; J.~Peoples,
  [Dark Energy Survey Collaboration],
  ``Dark Energy Survey''.
  
\bibitem{Guy:2016zel}
  J.~Guy [DESI Collaboration],
  ``The Dark Energy Spectroscopic Instrument''.

\bibitem{Tereno:2015hja}
  I.~Tereno {\it et al.} [Euclid Collaboration],
  IAU Symp.\  {\bf 306} (2014) 379
  doi:10.1017/S174392131401093X
  [arXiv:1502.00903 [astro-ph.IM]].

\bibitem{LSST}
  J.~R.~Lynne {\it et al.} [LSST collaboration],
  Icarus.\  {\bf 303} (2018),
  doi:10.1016/j.icarus.2017.11.033
  [arXiv:1711.10621 [astro-ph.EP]].

\bibitem{f}
 S. Nojiri, S. D. Odintsov,
 J. Phys. Conf. Ser. 66 012005 (2007),
[arxiv:hep-th/0611071]

\bibitem{8f}
 M. Chaichian, S. Nojiri, S. D. Odintsov, M. Oksanen, A. Tureanu,
 Class. Quant. Grav. 27 185021 (2010),
[arXiv:1001.4102]

\bibitem{PJ2007}
P. J. Zhang,
 Phys. Rev. D 76, 024007 (2007)
 
\bibitem{CLF2010}
 C. G. Boehmer, L. Hollenstein, F. S. N. Lobo and S. S. Seahra,
 Phys. Rev. D 76, 084005 (2007)

\bibitem{SS2003}
S. Nojiri, S. D. Odintsov,
 Phys. Rev. D 68, 123512 (2003)
[hep-th/0307288].

\bibitem{TFS2011}
 T. Harko, F. S. N Lobo, S. Nojiri, and S. D  Odintsov,
 Phys. Rev. D 84, 024020 (2011).

\bibitem{SSD2009}
S. Nojiri, S. D. Odintsov and D. S\'{a}ez-G\'{o}mez,
 Phys. Lett. B 681 74 (2009)
[arxiv:0908.1269].

\bibitem{ERV2010}
 E. Elizalde, R. Myrzakulov, V.V. Obukhov, D. Saez-Gomez,
 Class. Quantum Grav. 27 095007 (2010),
 [arXiv:1001.3636].

\bibitem{SS2011}
 S. Nojiri and S. D. Odintsov,
 Phys. Rept. 505, 59 (2011),
 [arXiv:1011.0544 [gr-qc]].

\bibitem{BI2007}
 B. M. Leith and I. P Neupane,
 J. Cosmol. Astropart. Phys. 0705, 019 (2007).

\bibitem{SSS2002}
 S. Nojiri, S. D. Odintsov and S. Ogushi,
 Int. J. Mod. Phys. A 17, 4809 (2002).
 
\bibitem{HD2013}
Hans-J\"{u}rgen Schmidt and Douglas Singleton,
 arXiv:1212.1769v2 [gr-qc]
 
\bibitem{g}
 S. Nojiri, S. D. Odintsov,
 Phys. Lett.B 631 (2005)
[arXiv:hep-th/0508049].

\bibitem{g1}
 G. Cognola, E. Elizalde, S. Nojiri, S. D. Odintsov, S. Zerbini,
 Phys. Rev. D 73(2006)084007,
[arXiv:hep-th/0601008].
 
\bibitem{g2}
S. Nojiri, S. D. Odintsov, M. Sasaki,
 Phys. Rev. D 71(2005)123509,
[arXiv:hep-th/0504052]

\bibitem{Becker:2017tcx}
  D.~Becker, C.~Ripken and F.~Saueressig,
  JHEP {\bf 1712} (2017) 121
  doi:10.1007/JHEP12(2017)121
  [arXiv:1709.09098 [hep-th]].
  
\bibitem{Motohashi:2014opa}
  H.~Motohashi and T.~Suyama,
  Phys.\ Rev.\ D {\bf 91} (2015) no.8,  085009
  doi:10.1103/PhysRevD.91.085009
  [arXiv:1411.3721 [physics.class-ph]].

\bibitem{Himmetoglu:2009qi}
  B.~Himmetoglu, C.~R.~Contaldi and M.~Peloso,
  Phys.~Rev.~D {\bf 80} (2009) 123530
  doi:10.1103/PhysRevD.80.123530
  [arXiv:0909.3524 [astro-ph.CO]].

\bibitem{KS}
H.~Kodama and M.~Sasaki,
Prog.~Theor.~Phys.~Suppl.~{\bf 78}, 1 (1984).

\bibitem{MFB} 
V.~F.~Mukhanov, H.~A.~Feldman and R.~H.~Brandenberger,
Phys.~Rept.~{\bf 215}, 203 (1992).

\bibitem{MM2008} 
K.~A.~Malik and D.~R.~Matravers,
Class.~Quant.~Grav.~{\bf 25}, 193001 (2008) 
[arXiv:0804.3276  [astro-ph]].

\bibitem{malik}
K.~A.~Malik and D.~Wands,
Phys. Rept.  {\bf 475}, 1 (2009)
[arXiv:0809.4944 [astro-ph]].

\bibitem{Bardeen80}
J.~M.~Bardeen,
Phys. Rev. D {\bf 22}, 1882 (1980).

\bibitem{FGM2} J.~L.~Fuentes, U.~A.~Gillani, and K.~A.~Malik
%
\emph{In preparation:} Vector and tensor perturbations in fourth-order
gravity (2019).

\bibitem{Abbott:2007kv}
  B.~P.~Abbott {\it et al.} [LIGO Scientific Collaboration],
  Rept. Prog. Phys.  {\bf 72} (2009) 076901,
  [arXiv:0711.3041 [gr-qc]].

\bibitem{Abbott:2009kk}
  B.~P.~Abbott {\it et al.} [VIRGO Collaboration],
  Astrophys. J.  {\bf 715} (2010) 1438,
  [arXiv:0908.3824 [astro-ph.HE]].

\bibitem{KAGRA}
  T.~Akutsu {\it et al.} [KAGRA Collaboration],
  Journal of Physics: Conference Series {\bf 610} (2015)

\bibitem{AmaroSeoane:2012je} 
  P.~Amaro-Seoane {\it et al.},
  Class.\ Quant.\ Grav.\  {\bf 29}, 124016 (2012)
  doi:10.1088/0264-9381/29/12/124016
  [arXiv:1202.0839 [gr-qc]].

\bibitem{xact}
	Jos\'{e} M. Mart\'{i}n-Garc\'{i}a,
    (2016),
    [http://www.xact.es/]
    
\bibitem{xpand}
  C.~Pitrou, X.~Roy and O.~Umeh,
  Class.~Quant.~Grav.~{\bf 30} (2013) 165002,
  doi:10.1088/0264-9381/30/16/165002,
  [arXiv:1302.6174 [astro-ph.CO]].
  
  \bibitem{nojiri:2018}
      S.~Nojiri, S.~D.~Odintsov and V.~K.~Oikonomou,
      ``{Ghost-free Gauss-Bonnet Theories of Gravity}",
      (2018),
      [arXiv:1811.07790 [gr-qc]].

\bibitem{githubGB}
	J.~L.~Fuentes,
	``Linear Gauss-Bonnet"
	(2018),
	[\texttt{https://github.com/jorchfv/Linear-Gauss-Bonnet}]












 











 



    

\end{thebibliography}
\end{document}